**Logarithmic thermodynamic cost of prediction from rare observations**

Andrey V. Dobrynin* [1] and Anastasia Stroujkova[2]

[1] *Department of Chemistry, University of North Carolina, Chapel Hill, NC, 27599, United States*

[2]*SeisLogic, LLC, Chapel Hill, NC, 27517, United States*

**Abstract:** A memory predicting changes in its environment from rare observations must spend energy on sampling and recording. We consider autonomous chemical memories driven by a sample-and-hold sensor observing an environment that switches at rate $\gamma$, with sampling frequency $f \ll \gamma$. An ideal observer with access to the full observation history can retain predictive information proportional to $f$. However, retaining a finite fraction of this information in an autonomous physical memory requires work of order $k_B T \ln(\gamma/f)$ per sensor change. This additional thermodynamic cost arises from the competing requirements of suppressing false records during long intervals between observations and responding rapidly to new observations. We establish a universal low-power bound on predictive information, which grows at most as $P/\ln\left(P_{ref}/P\right)$ with power budget $P$, uniformly over any finite number of memory states and independently of network topology and memory relaxation times. We further demonstrate that memory architecture determines the attainable scaling. With bounded transition rates and readout from the memory alone, two-state memories are limited to $P^2$ scaling, whereas a three-state memory is sufficient to attain the optimal logarithmic scaling. Our results establish a fundamental thermodynamic distinction between acquiring information and maintaining a useful record of a changing environment.

email: avd@email.unc.edu

## 1. Introduction.

Cells, sensors and engineered devices must anticipate changes in their environments using measurements that are noisy, infrequent and costly [1–7]. Understanding how such systems acquire and retain predictive information has motivated studies of information harvesting and dissipation in layered sensory memory systems [8], information acquisition and forgetting during sensory adaptation [9], statistical sufficiency and sensory efficiency [10], and predictive information and memory dissipation [11]. These studies build on the broader framework of information thermodynamics, which establishes the energetic costs associated with acquiring, storing and erasing information [12–18]. Adapting this framework to stationary prediction of an independently switching environment, we derive a universal bound on predictive information obtained from rare observations which is independent of the number of memory states. This addresses a fundamental question: Can a physical memory retain such information at an energetic cost proportional to the observation frequency, or does reliable recording require an additional thermodynamic penalty?

To distinguish the information available from observations from the thermodynamic cost of storing it, we first consider the information theoretic limit. Each observation carries a bounded amount of information, so the information gained can grow at most linearly with the sampling budget, as expressed by the capacity per unit cost [19]. We show that the Bayesian posterior constructed from the full observation record achieves this linear scaling. However, retaining this information in an autonomous chemical memory operating at finite power imposes an additional thermodynamic constraint that violates this scaling relation. For such a memory with $n$ internal states, coupled to a sample-and-hold sensor, we obtain that the maximum predictive information stored in the memory at power $P$ obeys

$$G_n^*(P) = \begin{cases} 0, & n = 1, \\ \Theta(P^2), & n = 2, \\ \Theta\left(P/\ln\left(P_{ref}/P\right)\right), & n \geq 3, \end{cases} \qquad P \to 0, \tag{1}$$

where $P_{ref}$ is a fixed power scale. For $n \geq 3$, the universal upper bound is independent of network topology and memory relaxation times, and remains valid when the decoder reads the sensor and memory jointly. The minimal three-state architecture that attains the logarithmic scaling is illustrated in Fig. 1(a). Its operation along with a representative stochastic trajectory is shown in Fig. 1(b).

The logarithm has a simple physical origin. A useful record of a rare event must be held in a rarely occupied state. Keeping that state rare in the background while allowing it to be reached rapidly following a sensor change requires an increasing energetic bias between the sensor-conditioned memory ensembles. The work needed to maintain this distinction grows logarithmically as observations become rare. This mechanism resembles kinetic proofreading [20–24], but with discrimination between recent and old events.

Thus, reliable prediction requires not only acquiring information about the environment but also paying a thermodynamic price for preserving its relevance over time.

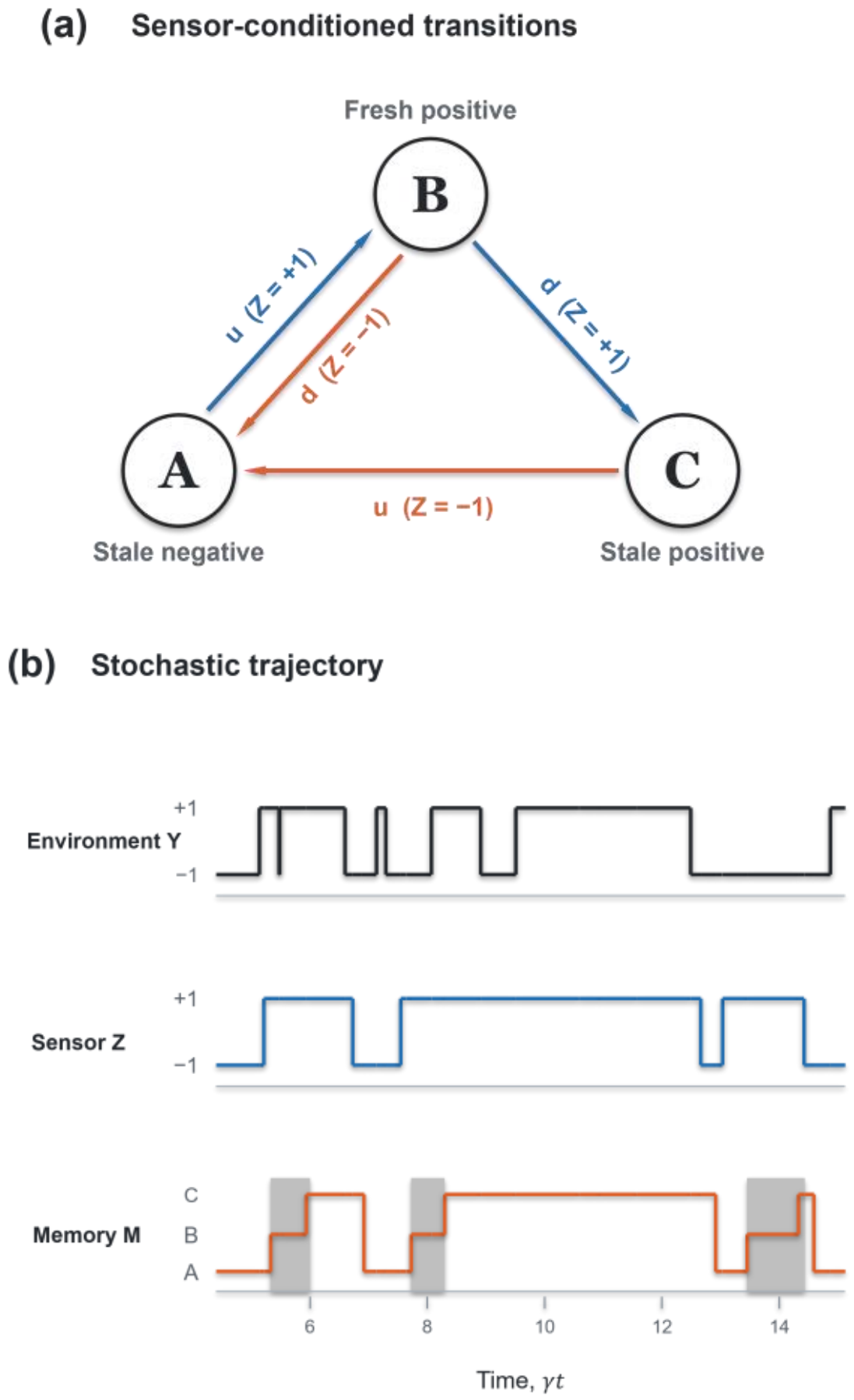

**FIG. 1. (a)** Minimal memory that attains the logarithmic order: stale-negative (A), fresh-positive (B) and stale-positive (C) states. Blue (orange) transitions are active when the held sensor value is Z = +1 (Z = −1); each has a reverse rate ε times smaller, with $\varepsilon \propto f$. **(b)** Simulated environment Y, held sensor Z with observation times (ticks), and memory state. Transient fresh state B (shaded) records recent negative-to-positive sensor changes. Illustrative parameters: $f = 2\gamma$, $\eta = 0.85$, $u = 7\gamma$, $d = 1.2\gamma$, $\varepsilon = 0.02$.

## 2. Model and Numerical Results

We consider a fluctuating environment described by a binary variable $Y = \pm 1$ that switches randomly between its two states at rate $\gamma$. Its temporal correlations decay as $\langle Y_t Y_{t+\tau} \rangle = e^{-2\gamma\tau}$, defining the characteristic timescale over which information about the environment remains useful. The environment is sampled at random times with average frequency $f$ [25]. Each observation provides a noisy measurement characterized by accuracy $0 < \eta \leq 1$ and requires energy $E_s$. The sensor retains its most recent reading $Z = \pm 1$ until the next observation [Fig. 1(b)]. The probability of obtaining a reading $z$ in environmental state $y$ is $\Pr(Z = z \mid Y = y) = (1 + \eta y z)/2$, giving a sensor switching rate $f(1 - \eta Y Z)/2$.

The sensor controls an autonomous memory with $n$ internal states, $M = 1, \ldots, n$. Transitions between memory states occur with rates $w_{ij}^z$ that depend on the held sensor value. We assume reversible transition channels with a maximum total escape rate κ. The sensor controls the transition rates without changing the internal state energies. Following local detailed balance [26], the power supplied to drive memory transitions is

$$P_m = k_B T \sum_z \sum_{i \neq j} a_z(i)\, w_{ij}^z \ln \frac{w_{ij}^z}{w_{ji}^z}, \qquad a_z(i) = \Pr(Z = z, M = i), \tag{2}$$

The total power budget is $P = f E_s + P_m$, accounting for both sampling and memory operation. Importantly, the memory responds to the held sensor value without direct access to observation times.

The memory performance is characterized by the predictive information $G_\tau = I(M_t; Y_{t+\tau})$ between the current memory state and the environment at a later time $\tau$. For a binary environment, this quantity can be expressed as

$$G_\tau = I(M_t; Y_{t+\tau}) = \sum_i p_i\, \Phi(q s_i); \ \Phi(x) = \frac{1 + x}{2} \ln(1 + x) + \frac{1 - x}{2} \ln(1 - x), \tag{3}$$

where $p_i = \Pr(M = i)$, $s_i = \langle Y_t \mid M_t = i \rangle$ and $q = e^{-2\gamma\tau}$. The parameter $s_i$ measures the correlation between memory state $i$ and the environment, while $q$ describes the loss of predictive information with increasing prediction time. Thus, useful prediction requires memory states that remain correlated with the environment.

We first establish the information theoretic limit by considering an ideal observer with access to the complete history of observations, including their times and outcomes. The Bayesian posterior provides the optimal estimate of the environment from this record. Between observations, the estimate loses information as environmental correlations decay, while each new measurement partially restores it. Balancing these processes in the stationary state gives an upper bound on predictive information (see Supporting Information):

$$G_\tau \leq \Phi(q) \frac{\eta^2 P}{4\gamma E_s + \eta^2 P}. \tag{4}$$

At low power, this bound scales linearly with $P$. The Bayesian posterior also achieves predictive information proportional to the observation frequency $f$ in the rare observation limit. Thus, an ideal observer can retain a finite fraction of the information provided by each measurement, even when observations are widely separated. However, an autonomous physical memory must retain this information through its internal dynamics, which introduces an additional thermodynamic constraint.

To understand this constraint, we examine how information about the environment is distributed among memory states. Since environmental correlations decay between observations, the total correlation retained by the memory is limited by the rate at which new measurements become available. This gives

$$\sum_i p_i |s_i| \le \frac{\eta f}{2\gamma}. \tag{5}$$

Equation (5) shows that the total environmental bias decreases linearly with the observation frequency. For weakly correlated states, Eq. (3) gives $\Phi(qs_i) \propto s_i^2$. Consequently, distributing the available correlation among commonly occupied states with weak environmental bias produces predictive information proportional to $f^2$. Retaining information proportional to $f$ instead requires rarely occupied states that remain strongly correlated with the environment.

These rare states provide a physical record of recent observations. During long intervals between observations, their background occupancy must remain small to prevent false records. Following a sensor change, however, they must become accessible before information about the environment is lost. As observations become rarer, satisfying both requirements demands an increasingly strong distinction between the memory distributions associated with different sensor readings. Maintaining this distinction against sensor driven mixing requires work that grows logarithmically with the inverse observation frequency.

The resulting thermodynamic constraint places a universal upper bound on predictive information. For sufficiently small power, we obtain

$$G_\tau \le C \, \frac{P}{k_B T \, \gamma \ln(2\gamma E_s / P)}, \qquad 0 < P < P_*. \tag{6}$$

Here the constants $C$ and $P_*$ depend on the physical parameters of the system but not on the number of memory states, network topology, or memory relaxation times. This bound also holds when predictive information is extracted jointly from the sensor and memory. Combining Eqs. (4) and (6) gives

$$G_\tau \le C' \, \frac{P}{\gamma [E_s + k_B T \ln(2\gamma E_s / P)]}. \tag{7}$$

Equation (7) reveals an additional thermodynamic penalty for recording rare observations. Beyond the sampling energy $E_s$, retaining a finite fraction of the available predictive information requires work of order $k_B T \ln(\gamma/f)$ per sensor change. This logarithmic penalty is independent of the number of memory states and cannot be eliminated by introducing more complex network topologies or slower relaxation modes.

The requirement for rare informative states places an important constraint on memory architecture. Consider a memory with only two internal states. If one state is rarely occupied, transitions into that state must be sufficiently slow to maintain its low background probability. However, this also prevents the memory from reaching that state rapidly after a sensor change. Alternatively, if both states are commonly occupied, their correlations with the environment become weak in the rare observation limit. In either case, the predictive information grows at most quadratically with the observation frequency, $G_\tau = O(f^2)$, for any two-state memory with bounded transition rates (Supplementary Material). Thus, two internal states cannot retain a finite fraction of the information available from rare observations when prediction is based on the memory state alone. This establishes the quadratic scaling for $n = 2$ summarized in Eq. (1). This restriction does not apply to joint sensor and memory readout, which can identify recent events through temporary differences between their states.

Adding a third state removes this limitation by allowing the memory to distinguish a recent observation from an older record. Figure 1(a) illustrates the simplest realization. States $A$ and $C$ represent previously recorded negative and positive sensor readings, while $B$ provides a short lived record of a recent positive observation. When the sensor is positive, the memory follows the sequence $A \to B \to C$, with transition rates $u$ and $d$. When the sensor is negative, the dynamics favor the return to $A$. Each transition has a reverse rate reduced by a factor $\varepsilon$, which controls the probability of spontaneous transitions in the opposite direction.

During long intervals between observations, the memory remains predominantly in $A$ or $C$. A negative to positive sensor change initiates a transition through $B$, which temporarily retains information about the environment [Fig. 1(b)]. To distinguish genuine recording events from spontaneous fluctuations, the background occupancy of $B$ must decrease with the observation frequency. Choosing $\varepsilon = cf/\gamma$, where $c$ is a constant, provides this separation while allowing the memory to respond within the environmental correlation time.

The resulting predictive information and memory power have the asymptotic forms

$$G_\tau = K_3 f[1 + o(1)]; \; P_m = \frac{3}{4} k_B T f \ln(1/\varepsilon)[1 + o(1)], \tag{8}$$

where $K_3 > 0$ depends on the transition rates, sensor accuracy, and prediction time. The memory therefore retains predictive information proportional to the observation frequency, while the work per sensor change increases logarithmically as observations become rarer.

The memory does not require continuous housekeeping dissipation while the sensor is held fixed. Each sensor conditioned transition network has no steady state cycling, and dissipation arises as the memory responds to changes in the sensor reading. The logarithmic cost originates from suppressing spontaneous transitions that would otherwise produce false records. This asymptotic behavior requires the reverse rates to decrease with the observation frequency rather than remaining fixed.

Combining the two relations in Eq. (8) with the total power budget gives $G_\tau = \Theta\left[P/\ln\left(P_{ref}/P\right)\right]$. Together with the universal upper bound in Eq. (7), this construction establishes the logarithmic scaling for $n \geq 3$ summarized in Eq. (1). Thus, three internal states are sufficient to achieve the optimal asymptotic scaling, and increasing the number of states cannot improve its dependence on power.

Figure 2 compares the low-power prediction–power behavior of two-, three-, and four-state memories. The exact two-state result decreases quadratically with power, $G_\tau \sim P^2$, and therefore falls rapidly below the information available from the observation history. In contrast, the three-state construction retains substantially more predictive information and approaches the logarithmic scaling $G_\tau \sim P / \ln(P_{ref}/P)$ derived above (Eq. 7). An optimized, sign-symmetric four-state design, with no prescribed leakage law, lies above the three-state reference at the plotted powers and is consistent with the same favorable scaling. These are feasible restricted topology results, not a certified global optimum.

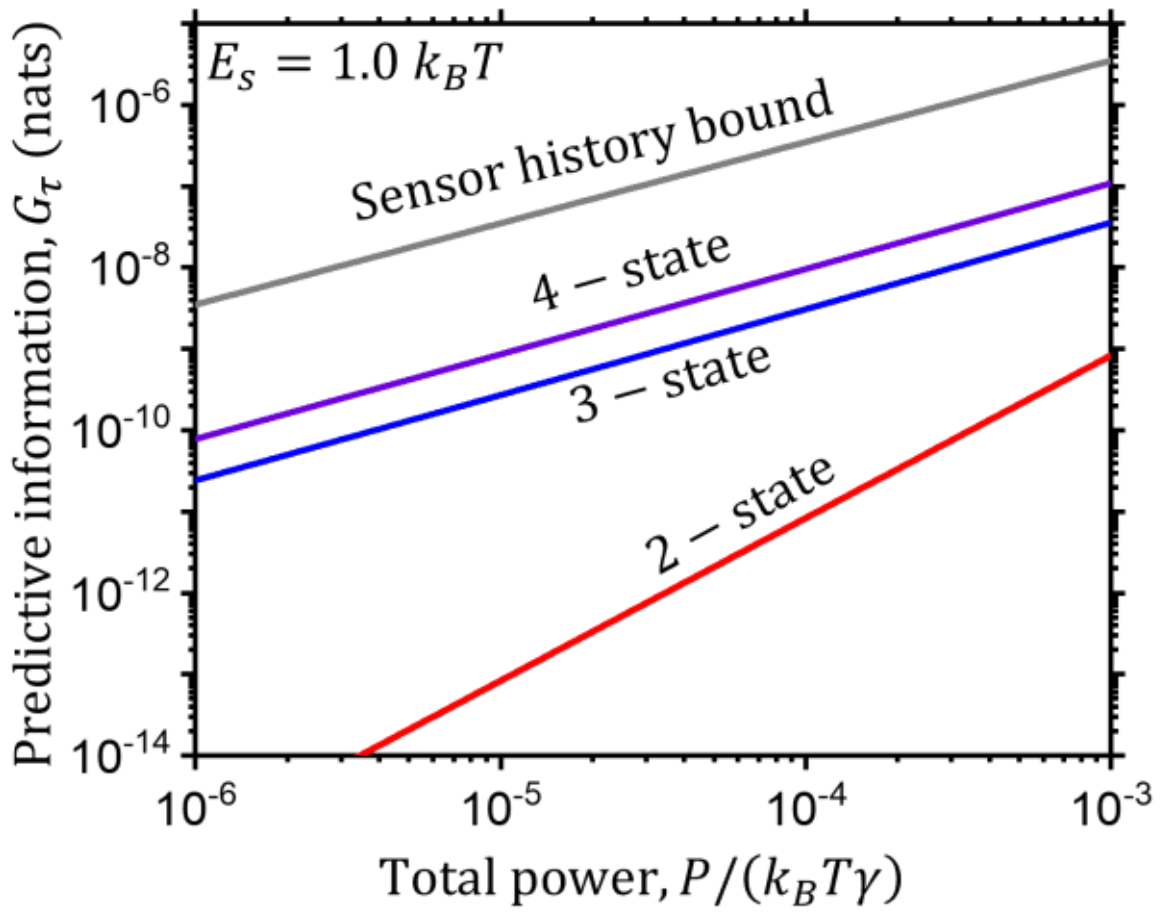

**FIG. 2.** Predictive information versus total power budget ($\gamma = 1$, $k_B T = 1$, $\eta = 0.5$, $\gamma\tau = 0.4$, $\kappa = 10\gamma$, and $E_s = 1.0\ k_B T$). The gray line is the sensor history bound which scales linearly with power $P$. The optimized two-state reactive memory result (red line) scales as $G_\tau \sim P^2$, and the three-state memory construction (blue line) attains the optimal logarithmic scaling $G_\tau \sim P / \ln(P_{ref}/P)$. Purple line corresponds to sign-symmetric four-state design with six rates and $f$ optimized at each power budget (no prescribed leakage, numerical trend only). For three-state construction: $u = 7\gamma$, $d = 3\gamma$, $c = 0.01$. Calculation details are provided in Supplementary Material.

Figure 3 connects the information retained per rare observation to the logarithmically increasing work required to record it. For the three-state construction, $\gamma G_\tau / f$ approaches a nonzero constant as $f/\gamma \to 0$ [Fig. 3(a)], demonstrating that a finite amount of predictive information is retained per rare observation.

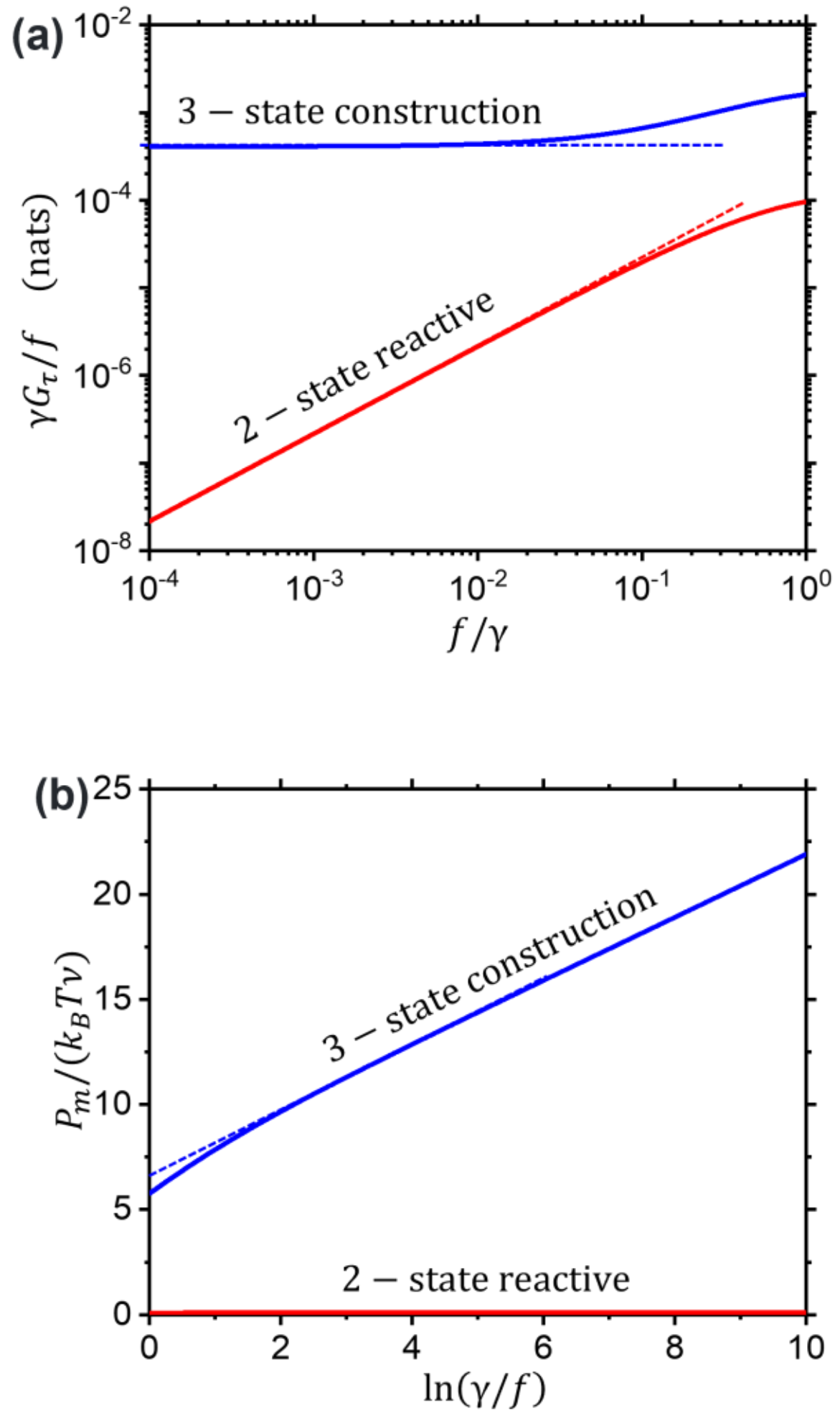

**FIG. 3.** Information retained and energetic cost in the rare observation limit for the three-state construction (blue line) and two-state reactive memory (red line). **(a)** Predictive information normalized by the observation frequency, $\gamma G_\tau/f$, versus $f/\gamma$. For the three-state construction, $\gamma G_\tau/f$ approaches a finite constant for $f/\gamma \to 0$, corresponding to $G_\tau \sim f$. The two-state result vanishes, consistent with $G_\tau \sim f^2$. **(b)** Normalized memory work per sensor change, $P_m/k_B T\nu$ ($\nu$ is the mean sensor-change rate), as a function of $\ln(\gamma/f)$. The three-state construction shows the predicted logarithmic increase in recording work as observations become rare, while the two-state memory has a much smaller energetic cost but retains asymptotically negligible information. Dotted lines indicate the corresponding asymptotic results. Parameters are the same as in Fig. 2.

The corresponding two-state quantity vanishes, as expected from $G_\tau = O(f^2)$. Achieving the linear information scaling requires an increasing recording cost: the normalized memory work per sensor change grows as $\ln(\gamma/f)$ for the three-state construction [Fig. 3(b)]. The two-state memory dissipates much less work, but this apparent energetic advantage comes at the cost of losing essentially all predictive information per observation in the rare observation limit.

The results apply to autonomous memories controlled by the held sensor value, without direct access to observation times or an external clock. No assumption about relaxation between observations is required, so slow internal dynamics and metastability cannot eliminate the logarithmic penalty. When sensor changes modify internal state energies, the associated work must also be included in the energetic cost. For passive sensors with negligible sampling cost, the power dependent bound in Eq. (7) is replaced by the corresponding constraint on work per sensor change, which retains the logarithmic penalty for memories capturing a finite fraction of the available information (Supplementary Material). The quadratic bound for two-state memories requires bounded transition rates, whereas the universal logarithmic bound requires this restriction only for perfectly accurate sensors.

## 3. Conclusions.

Our results identify a fundamental difference between acquiring information and maintaining a useful record in an autonomous physical system. An ideal observer can extract predictive information proportional to the observation frequency, but a physical memory must also distinguish recent observations from the much more common background of outdated records. This requirement introduces an additional thermodynamic penalty that cannot be eliminated by increasing the number of memory states or changing network topology.

The relative importance of sampling and recording costs depends on the observation frequency. When measurements are energetically expensive, sampling dominates except at extremely low frequencies. For inexpensive or passive measurements, however, the energetic cost of reliable recording can become the main limitation. This distinction is particularly relevant to molecular sensing and biochemical signaling networks, where observations may arise from spontaneous binding events while memory formation requires driven molecular transitions.

The logarithmic penalty differs from the finite thermodynamic cost of erasing a stored bit [12,27]. It originates from the competing requirements of suppressing false records during long waiting periods and responding rapidly to new observations. The three-state construction demonstrates how these requirements can be satisfied using a transient recording state followed by active resetting. Similar mechanisms may operate in sensing networks containing short lived molecular modification states.

The fundamental limitation is therefore not the amount of information carried by a rare observation, but the thermodynamic cost of distinguishing a fresh record from an outdated one.

**Supplementary Material**

Details of the model and thermodynamic work accounting, derivations of the information bounds for ideal observers and autonomous memories, stationary state equations, entropy balance, the universal logarithmic work bound, the analysis of two-state and three-state memory architectures, and additional numerical calculations.

**Data Availability**

The data are available within the article and its supplementary material.

# Supplementary Material

## Logarithmic thermodynamic cost of prediction from rare observations

Andrey V. Dobrynin*[1] and Anastasia Stroujkova[2]
[1] *Department of Chemistry, University of North Carolina, Chapel Hill, NC, 27599, United States*
[2] *SeisLogic, LLC, Chapel Hill, NC, 27517, United States*
email: avd@email.unc.edu

### S1. Model, dynamics, and work accounting

**Physical setting.** The environment $Y = \pm 1$ switches independently at rate $\gamma$. Observations occur at Poisson rate $f$ and overwrite the sample-and-hold sensor $Z = \pm 1$ with a noisy reading of accuracy $\eta$. The autonomous memory has $n$ internal states $M = 1, \dots, n$, and its transition rates depend only on the currently held value $Z$. The allowed changes of the joint state $(Y, Z, M)$ are

$$Y \to -Y\colon\ \gamma; \qquad Z \to -Z\colon\ \frac{f}{2}(1 - \eta Y Z); \qquad i \to j\colon\ w_{ij}^{z}. \tag{S1.1}$$

The sensor transition rate follows from the probability $(1 + \eta Y z)/2$ of a reading $z$. A new observation that leaves $Z$ unchanged is invisible to the memory, although it still consumes $E_s$. Memory transitions are not synchronized to sampling events and the memory receives no clock or timestamp.

For each sensor value $z$, the generator $W_z$ has off diagonal elements $w_{ij}^{z}$. Its diagonal entries and the bound on the total exit rate are

$$(W_z)_{ii} = -\sum_{j \neq i} w_{ij}^{z}, \qquad \sum_{j \neq i} w_{ij}^{z} \leq \kappa. \tag{S1.2}$$

Every active transition channel has a reverse channel. The intrinsic memory energies are equal and independent of the sensor; $Z$ gates chemical channels rather than changing energy levels. By local detailed balance, the chemical work associated with channel $i \to j$ at fixed $z$ is $k_B T \ln\left(w_{ij}^{z}/w_{ji}^{z}\right)$. Its stationary rate is

$$P_m = k_B T \sum_{z} \sum_{i \neq j} a_z(i) w_{ij}^{z} \ln\left(\frac{w_{ij}^{z}}{w_{ji}^{z}}\right), \qquad a_z(i) = \Pr(Z = z, M = i). \tag{S1.3}$$

If parallel physical channels connect the same two states, the work is evaluated channel by channel. The log sum inequality shows that this is at least the aggregate Eq. (S1.3), so the lower bounds remain valid. The budget is $P \geq f E_s + P_m$, with $E_s > 0$ in the power-dependent theorem. Section S10 establishes $P_m \geq 0$ for the stipulated chemical implementation; hence $f \leq P/E_s$. We consider stationary causal realizations with a unique stationary distribution and memory noise independent of $Y$ once the sensor history is fixed. Symmetry gives $\Pr(Z = +1) = \Pr(Z = -1) = 1/2$. Whenever a subscript names a sensor state, $+1$ and $-1$ refer to its two held readings.

**Predictive information.** For memory state $i$, we define its stationary probability, environmental correlation, and conditional environmental bias as

$$p_i = \Pr(M = i), \qquad c_i = \langle Y\mathbf{1}_{M=i}\rangle, \qquad s_i = \frac{c_i}{p_i} = \langle Y \mid M = i\rangle. \tag{S1.4}$$

The environment propagator is

$$\Pr(Y_{t+\tau} = y' \mid Y_t = y) = \frac{1 + qyy'}{2}, \tag{S1.5}$$

where $q = exp(-\lambda\tau)$ with $\lambda = 2\gamma$.

The predictive information extracted from the memory alone is

$$G_\tau = I(M_t; Y_{t+\tau}) = \sum_i p_i\, \Phi(qs_i), \tag{S1.6}$$

with

$$\Phi(x) = \frac{1 + x}{2}\ln(1 + x) + \frac{1 - x}{2}\ln(1 - x). \tag{S1.7}$$

The power series

$$\Phi(x) = \sum_{k\geq 1} \frac{x^{2k}}{2k(2k - 1)} \tag{S1.8}$$

has nonnegative coefficients and implies $\Phi(qs) \leq \Phi(q)s^2$ for $|s| \leq 1$. Thus

$$G_\tau \leq \Phi(q)\sum_i p_i\, s_i^2 = \Phi(q)\sum_i |c_i|\,|s_i|. \tag{S1.9}$$

## S2. Information available from observations

### S2.1. Ideal observer with the complete measurement record

The ideal observer knows all observation times and outcomes. Its Bayesian posterior mean $h = \langle\, Y \mid \text{observation history}\,\rangle$ evolves between observations according to $\dot{h} = -\lambda h$. A measurement with outcome $z$ updates the posterior according to Bayes' rule:

$$h \mapsto \frac{h + \eta z}{1 + \eta z h}. \tag{S2.1}$$

The conditional mean of this update equals $h$, and its variance is

$$\mathrm{Var}\big(h_{\text{post}} \mid h\big) = \frac{\eta^2(1 - h^2)^2}{1 - \eta^2 h^2}. \tag{S2.2}$$

The stationary second moment therefore balances information gained from observations against its decay through environmental switching:

$$2\lambda\langle h^2\rangle = f\eta^2\left\langle\frac{(1 - h^2)^2}{1 - \eta^2 h^2}\right\rangle \leq f\eta^2(1 - \langle h^2\rangle). \tag{S2.3}$$

This inequality uses $1-h^2 \le 1-\eta^2 h^2$ for $0 \le \eta \le 1$. It follows that $\langle h^2\rangle \le f\eta^2/(2\lambda + f\eta^2)$. Since the physical memory is a causal representation of the record, data processing gives $G_\tau \le \langle \Phi(qh)\rangle \le \Phi(q)\langle h^2\rangle$. With the sampling budget $fE_s \le P$,

$$G_\tau \le \Phi(q)\frac{f\eta^2}{4\gamma + f\eta^2} \le \Phi(q)\frac{\eta^2 P}{4\gamma E_s + \eta^2 P}. \tag{S2.4}$$

The last expression is Eq. (4) of the main text. It is linear in $P$ for small budgets. The posterior attains this linear scaling in the rare observation limit. When $f/\lambda \to 0$, measurements are isolated on the environmental correlation time. Following a single observation, $h \simeq \eta z e^{-\lambda t}$ until the next observation. Averaging over the age of the latest observation yields the predictive information available from the complete observation history $\mathcal{H}_t$:

$$U_\tau(f) \equiv I(\mathcal{H}_t; Y_{t+\tau}) = \frac{f}{\lambda}\int_0^1 \frac{\Phi(q\eta s)}{s}\, ds + o(f). \tag{S2.5}$$

The coefficient of $f$ is finite and strictly positive when $\eta > 0$ and $q > 0$. Thus $U_\tau(f) = \Theta(f)$ for rare observations. The record includes observations that do not change $Z$, so this result does not imply that a downstream autonomous memory can capture the same information.

As a numerical benchmark, for $\gamma = 1, \eta = 0.5$, and $\tau = 0.4$, direct evaluation of the stationary posterior gives 0.0106, 0.0343, and 0.0862 nats at $f = 2$, 10, and 150. The corresponding first bound in Eq. (S2.4) is 0.0116, 0.0403, and 0.0946 nats, within 20% of the exact values.

### S2.2. Observer restricted to sensor changes

An autonomous memory driven by $Z$ has access to its history of changes, not to all observations. Let $h_Z(t) = \langle Y_t \mid Z_{[0,t]}\rangle$. The absence of a sensor change also carries information: a long hold makes agreement between $Z$ and $Y$ more likely. Point process filtering [1] gives, between changes,

$$\dot{h}_Z = -\lambda h_Z + \frac{f\eta}{2} Z(1-h_Z^2). \tag{S2.6}$$

At a sensor change, the posterior jumps according to

$$h_Z \mapsto \frac{h_Z - \eta Z}{1 - \eta Z h_Z}, \tag{S2.7}$$

where $Z$ is the value immediately before the change. The stationary second moment satisfies

$$2\lambda\langle h_Z^2\rangle = \frac{f\eta^2}{2}\left\langle\frac{(1-h_Z^2)^2}{1-\eta Z h_Z}\right\rangle. \tag{S2.8}$$

The function $(1-m^2)/(1-\eta m)$ has maximum $c_\eta = 2/\left[1+\sqrt{1-\eta^2}\right]$ on $|m| \le 1$. The corresponding bound is

$$G_\tau \le \Phi(q)\frac{c_\eta f\eta^2}{8\gamma + c_\eta f\eta^2} \le \Phi(q)\frac{c_\eta \eta^2 P}{8\gamma E_s + c_\eta \eta^2 P}. \tag{S2.9}$$

This is the sensory benchmark plotted in Figure 2 of the main text. Its rare sampling coefficient is lower than the complete record benchmark because the memory cannot detect observations that leave $Z$ unchanged. Event driven simulations of the filter give $0.0064$, $0.0239$, and $0.0762$ nats at $f = 2$, $10$, and $150$ for the parameters above.

## S3. Why informative memory states must be rare

**Bias budget.** Any memory that obtains information through the observations inherits a limit on its total instantaneous correlation with the environment. This limit holds for an arbitrary causal readout $X$, including $M$ alone or $(Z, M)$ jointly:

$$\sum_x |\langle Y \mathbf{1}_{X=x} \rangle| \leq \langle |h| \rangle \leq \frac{\eta f}{\lambda}. \tag{S3.1}$$

To prove the second inequality, average $|h|$ just after an observation over its outcome. Averaging the update over observation outcomes gives

$$\langle |h_{\text{post}}| \mid h_{\text{pre}} \rangle = \max(|h_{\text{pre}}|, \eta). \tag{S3.2}$$

Between observations $|h|$ decays at rate $\lambda$. Stationarity then requires

$$\lambda \langle |h| \rangle = f \langle \max(\eta - |h|, 0) \rangle \leq f\eta. \tag{S3.3}$$

The first inequality follows from $\langle Y \mid X \rangle = \langle h \mid X \rangle$ and Jensen's inequality. Equation (S3.1) is Eq. (5) of the main text for $X = M$.

**Rare and strongly correlated states.** Suppose $G_\tau \geq af$ along a sequence $f \to 0$ for some fixed $a > 0$. We define the set of informative states $A_b = \{i \colon |s_i| \geq b\}$ with $b = a\lambda/[2\Phi(q)\eta]$ (for a nontrivial benchmark). According to Eq. (S1.9), the contribution from the complement is at most $\Phi(q) b \sum_i |c_i| \leq af/2$. Hence at least $af/2$ of the predictive information lies in $A_b$, and $p(A_b) \geq af/[2\Phi(q)]$. Conversely, Eq. (S3.1) gives $bp(A_b) \leq \eta f/\lambda$. We conclude

$$p(A_b) = \Theta(f), \qquad G_\tau \geq af > 0 \text{ with fixed } a \text{ as } f \to 0. \tag{S3.4}$$

This argument formalizes the rare state picture used in the main text: preserving information proportional to $f$ demands a set that is infrequently occupied yet maintains a finite environmental bias. If the number of internal states grows with decreasing $f$, the informative probability can be spread over many states. By contrast, information spread among commonly occupied states with biases of order $f$ is only $O(f^2)$.

## S4. Stationary equations for sensor and memory correlations

To obtain a bound that applies even when some memory modes are arbitrarily slow, we work with the actual stationary sensor-conditioned distributions. The row vectors are defined $a_z(i) = \Pr(Z = z, M = i)$ and $b_z(i) = \langle Y \mathbf{1}_{Z=z, M=i} \rangle$, with $p = a_{+1} + a_{-1}$ and $c = b_{+1} + b_{-1}$. The sensor flip flux out of $(z, i)$ is

$$j_z(i) = \frac{f}{2}[a_z(i) - \eta z b_z(i)]. \tag{S4.1}$$

Summing the full master equation over $Y$, and then weighting it by $Y$ before summation, gives the exact stationary relations

$$0 = a_z W_z + \frac{f}{2}(a_{-z} - a_z) + \frac{f\eta z}{2} c, \tag{S4.2}$$

$$0 = b_z W_z - \lambda b_z + \frac{f}{2}(b_{-z} - b_z) + \frac{f\eta z}{2} p. \tag{S4.3}$$

Applying Eq. (S3.1) to the joint readout $X = (Z, M)$ gives $\| b_{+1} \|_1 + \| b_{-1} \|_1 \leq \eta x$ with $x = f/\lambda$.

## S5. Entropy balance and the work needed to restore memory distinctions

A sensor change redistributes probability between the two memory ensembles conditioned on $Z = \pm 1$. Memory transitions must counteract this mixing if the ensembles are to remain distinguishable. The sensor-conditioned memory ensembles are $r_z = 2a_z = \Pr(M = \cdot \mid Z = z)$. Their symmetric relative entropy is

$$J = D(r_{+1} \| r_{-1}) + D(r_{-1} \| r_{+1}). \tag{S5.1}$$

The quantity $J$ measures how differently the memory is distributed under the two sensor readings. We define the two parameters

$$\rho_f = \frac{\kappa + f}{\kappa + f + \lambda}, \qquad \delta_f = 1 - \eta \rho_f > 0. \tag{S5.2}$$

Stationarity and nonnegative memory entropy production then imply

$$P_m \geq k_B T f \left( \frac{\delta_f J}{4} - \frac{\eta \rho_f}{e} \right). \tag{S5.3}$$

The entropy balance follows from the Shannon entropy

$$H = -\sum_{z,i} a_z(i) \ln a_z(i). \tag{S5.4}$$

Stationarity requires $\dot{H}_M + \dot{H}_Z = 0$, where the two terms describe memory jumps and sensor changes. The coarse-grained entropy production associated with memory transitions is

$$\Sigma_M = \sum_z \sum_{i \neq j} a_z(i) w_{ij}^z \ln \left[ \frac{a_z(i) w_{ij}^z}{a_z(j) w_{ji}^z} \right] \geq 0. \tag{S5.5}$$

The nonnegativity follows by pairing forward and reverse fluxes. Comparing the stationary work and entropy balances gives

$$\frac{P_m}{k_B T} = \Sigma_M - \dot{H}_M = \Sigma_M + \dot{H}_Z \geq \dot{H}_Z. \tag{S5.6}$$

The entropy change produced by sensor flips is

$$\dot{H}_Z = \frac{f}{2} \sum_i [a_{+1}(i) - a_{-1}(i) - \eta c_i] \ln \frac{a_{+1}(i)}{a_{-1}(i)}. \tag{S5.7}$$

Its first part equals $fJ/4$. To bound the term involving environmental correlations, note that stationarity at fixed $(y, z, i)$ implies $\gamma \Pr(-y, z, i) \leq (\gamma + \kappa + f) \Pr(y, z, i)$. Interchanging $y$ and $-y$ yields $|b_z(i)| \leq \rho_f a_z(i)$ and $|c_i| \leq \rho_f p_i$. Using $\min(u, v) |\ln(u/v)| \leq \max(u, v)/e$ bounds the remaining contribution from below by $-f\eta\rho_f(J/4 + 1/e)$ and gives Eq. (S5.3). For imperfect sensors ($\eta < 1$), $|c_i| \leq p_i$ can be used in place of the rate-dependent estimate, removing the need for the exit cap. For a single channel between each pair and equal intrinsic energies, the entropy identity is equality.

## S6. Why slow relaxation does not invalidate the bound

A conventional argument would replace the actual conditional memory distribution by the stationary distribution of the frozen generator $W_z$. Instead, we examine its response on the environmental correlation time $1/\lambda$. To do so, we define the resolvent and its normalized propagation kernel:

$$R_z = (\lambda - W_z)^{-1}, \qquad K_z = \lambda R_z. \tag{S6.1}$$

The kernel $K_z$ propagates the frozen memory for an exponentially distributed time of mean $1/\lambda$ and remains well defined even for reducible $W_z$. For any row vector $v$, $\| vR_z \|_1 \leq \| v \|_1/\lambda$.

Solving Eq. (S4.3) for $b_z$ and summing over sensor states gives

$$c = \frac{\eta x}{2} p(K_{+1} - K_{-1}) + e_f, \qquad \| e_f \|_1 \leq \eta x^2, \tag{S6.2}$$

where $x = f/\lambda$. Equation (S4.2), together with the bias budget, gives $\| r_z W_z \|_1 \leq 2f$ for $f \leq \lambda$. Because $K_z - I = W_z R_z$,

$$\| r_z K_z - r_z \|_1 \leq 2x. \tag{S6.3}$$

Stationarity and rare sampling imply that the conditional memory ensemble changes little over one environmental correlation time under the corresponding frozen sensor dynamics. This result requires no assumption of relaxation to a frozen stationary distribution, making the final bound independent of memory relaxation times.

## S7. Finite sampling bound and logarithmic work penalty

### S7.1. Rare sets require separation of conditional memory distributions

Introduce the observation rarity parameter and two auxiliary quantities:

$$L = \ln(1/x), \qquad B = 2x(\eta L + 1), \qquad \ell = \ln(1/B). \tag{S7.1}$$

For $L > 1$ and $B < 1$, the finite sampling estimate is

$$G_\tau \leq \Phi(q)\eta x \left[\frac{1}{L} + \frac{B}{2} + \frac{J + 2\ln 2}{4\ell} + x\right]. \tag{S7.2}$$

To understand the estimate, consider the potentially informative set $A = \{i : |s_i| > 1/L\}$. Equation (S3.1) gives $p(A) \leq \eta x L$, while Eq. (S6.3) shows that this set remains rare under dynamics conditioned on the same sensor value: $(r_z K_z)(A) \leq B$. Now consider an ensemble prepared under the opposite sensor reading, and let $\alpha_z = (r_{-z} K_z)(A)$ denote its probability of entering $A$ within an environmental correlation time after the held value becomes $z$. Relative entropy cannot increase under the same stochastic kernel $K_z$; coarse graining into $A$ and its complement therefore gives

$$D(r_{-z} \| r_z) \geq D(r_{-z} K_z \| r_z K_z) \geq \alpha_z \ln(1/B) - \ln 2. \tag{S7.3}$$

The transition probabilities satisfy

$$\sum_z \alpha_z \le \frac{J + 2\ln 2}{\ell}, \qquad \sum_z (pK_z)\,(A) \le B + \frac{J + 2\ln 2}{2\ell}. \tag{S7.4}$$

Inserting these inequalities into Eq. (S6.2) bounds the environmental correlation carried by $A$:

$$\sum_{i \in A} |c_i| \le \frac{\eta x}{2}\left[B + \frac{J + 2\ln 2}{2\ell}\right] + \eta x^2. \tag{S7.5}$$

Outside $A$, the contribution to Eq. (S1.9) obeys $p_i s_i^2 \le |c_i|/L$. Inside $A$, it is bounded by $|c_i|$. Combining the two contributions proves Eq. (S7.2). An autonomous memory must make rare informative states accessible after a sensor change while suppressing their background occupation. Accomplishing both requires a separation of the two conditional ensembles that grows logarithmically with $1/f$.

**S7.2. Universal power bound**

We define the constants

$$\rho_\star = \frac{\kappa + \lambda}{\kappa + 2\lambda}, \qquad \delta_\star = 1 - \eta\rho_\star > 0, \tag{S7.6}$$

$$A_\star = 2 + \ln 2 + \frac{2\eta\rho_\star}{e\delta_\star}, \qquad P_\star = \lambda E_s e^{-10}. \tag{S7.7}$$

For $0 < P < P_\star$, every finite-state autonomous memory obeys

$$G_\tau \le C \frac{P}{\ln(\lambda E_s/P)}, \qquad C = \frac{\Phi(q)\eta}{\lambda} \max\left(\frac{A_\star}{E_s}, \frac{2}{\delta_\star k_B T}\right). \tag{S7.8}$$

For $L \ge 10$, $B \le 2e^{-L}(L+1)$, so $\ell \ge L/2$ and $B/2 + x \le 1/L$. Equation (S7.2) then reduces to

$$G_\tau \le \frac{\Phi(q)\eta f}{\lambda L}\left(2 + \ln 2 + \frac{J}{2}\right). \tag{S7.9}$$

For $f \le \lambda$, Eq. (S5.3), together with $\rho_f \le \rho_\star$ and $\delta_f \ge \delta_\star$, gives

$$\frac{J}{2} \le \frac{2P_m}{k_B T \delta_\star f} + \frac{2\eta\rho_\star}{e\delta_\star}. \tag{S7.10}$$

Combining the two inequalities yields

$$G_\tau \le \frac{\Phi(q)\eta}{\lambda L}\left(A_\star f + \frac{2P_m}{k_B T \delta_\star}\right). \tag{S7.11}$$

The total budget satisfies

$$A_\star f + \frac{2P_m}{k_B T \delta_\star} \le \max\left\{\frac{A_\star}{E_s}, \frac{2}{k_B T \delta_\star}\right\}(f E_s + P_m). \tag{S7.12}$$

Together with $f \le P/E_s$, this implies $L \ge \ln(\lambda E_s/P)$ and establishes Eq. (S7.8), corresponding to Eq. (6) in the main text.

**Uniformity and combined cost.** The bound holds for any finite number of memory states $n = n(P)$, transition graph, and relaxation spectrum, even when the rates are retuned as $P$ changes. Combining Eqs. (S2.4) and (S7.8) gives Eq. (7) of the main text, up to a numerical prefactor.

**Joint sensor and memory readout.** The same logarithmic power dependence holds if the decoder observes $(Z, M)$ rather than $M$ alone. The bias budget (S3.1) holds for the joint state, and the stationary equation (S4.3) supplies the required conditioned correlation vectors. Repeating the rare set construction with sets of pairs $(z, i)$ and using the same entropy balance produces Eq. (S7.8) with changed constants but unchanged scaling. This extension is important because only the stronger quadratic restriction for $n = 2$ relies on readout from $M$ alone.

### S7.3. Explicit work per sensor change

Although the power formulation assumes $E_s > 0$, the work-per-change bound survives the passive sensor limit $E_s \to 0$. Assuming that the memory captures a fixed fraction $\theta > 0$ of the linear sensory scale, we have

$$G_\tau \geq \theta \Phi(q) \frac{\eta^2 f}{2\lambda}. \tag{S7.13}$$

The simplified rare sampling bound requires

$$\frac{J}{2} \geq \frac{\theta \eta L}{2} - (2 + \ln 2). \tag{S7.14}$$

Combining this with the entropy balance in Eq. (S5.3) gives, for $f \leq \lambda e^{-10}$,

$$\frac{P_m}{k_B T \nu} \geq \frac{\theta \eta \delta_\star}{2} \ln\left(\frac{\lambda}{f}\right) - \delta_\star A_\star, \tag{S7.15}$$

where $\nu \leq f/2$ is the stationary rate of sensor changes. This bound gives a quantitative form of the logarithmic work penalty discussed in the main text. It shows that memory work per useful sensor change grows as $k_B T \ln(\lambda/f)$ whenever a fixed fraction of the available rare observation information is retained.

## S8. Why two memory states give quadratic information

With only two internal states, a state that is rare in the background cannot also be entered promptly after a sensor change while the exit rates remain capped. This observation can be turned into a uniform bound for arbitrary sensor dependent two state transition rates:

$$G_\tau \leq \Phi(q) C_2^2 f^2, \qquad C_2 = \frac{\eta}{\lambda} \max\left\{4, 1 + \frac{4\kappa}{\lambda}\right\}. \tag{S8.1}$$

Let $u_z$ and $v_z$ be the rates of transitions $0 \to 1$ and $1 \to 0$, respectively. If $p = \Pr(M = 1) \leq 1/4$, stationarity of the marginal memory population gives

$$\sum_z a_z(0) u_z = \sum_z a_z(1) v_z \leq \kappa p. \tag{S8.2}$$

Because $a_z(0) = 1/2 - a_z(1) \geq 1/4$, both entrance rates satisfy $u_z \leq 4\kappa p$. Thus, the small steady population of state 1 enforces small entrance rates under both sensor values.

The generator $L_0$ for $(Z, M)$ retains actual memory transition rates but replaces sensor changes with independent flips at rate $f/2$. Equation (S4.3), written for the two sensor blocks, has a source proportional to $(-p_M, p_M)$, where $p_M = (1 - p, p)$. Its resolvent representation gives

$$c_1 = \frac{f\eta}{2}\int_0^\infty e^{-\lambda t}\left[\Pr_{+1}(M_t = 1) - \Pr_{-1}(M_t = 1)\right] dt. \tag{S8.3}$$

The two probabilities start with the same memory distribution $p_M$, but opposite held sensor values. Since each entrance rate is at most $4\kappa p$, either occupation probability is bounded by $p + 4\kappa p t$. Thus,

$$|c_1| \leq \frac{f\eta p}{2}\left(\frac{1}{\lambda} + \frac{4\kappa}{\lambda^2}\right). \tag{S8.4}$$

Both conditional biases $|s_0|$ and $|s_1|$ are bounded by $C_2 f$. For $1/4 \leq p \leq 3/4$, the bias budget (S3.1) gives $|s_i| \leq 4\eta f/\lambda$. The case $p \geq 3/4$ follows by interchanging the state labels. Substitution into Eq. (S1.9) proves Eq. (S8.1), establishing the upper half of the $n = 2$ case in Eq. (1) of the main text.

**An explicit quadratic construction.** The quadratic scaling is attained by a reactive two-state memory. For this construction the two memory outputs are encoded as $M = \pm 1$. The memory has fixed positive transition rates $u$ and $v$, with $r = u + v \leq \kappa$, affinity $F = \ln(u/v) > 0$, and $c_F = \tanh(F/2)$. When the favored direction reverses with $Z$, exact moment balance gives the stationary environment memory correlation $b = \langle YM\rangle$ and memory power

$$b = \frac{\eta f}{f+\lambda}\frac{r}{r+\lambda}c_F = \frac{\eta r c_F}{\lambda(r+\lambda)}f + O(f^2), \tag{S8.5}$$

$$P_m = \frac{k_B T}{2} f F c_F + O(f^2). \tag{S8.6}$$

Writing $A = E_s + (k_B T/2)F c_F$, we have $P = Af + O(f^2)$ and $G_\tau = \Phi(qb) = q^2 b^2/2 + O(b^4)$. Therefore

$$G_\tau = \frac{q^2\eta^2 r^2 c_F^2}{2\lambda^2(r+\lambda)^2 A^2}P^2 + o(P^2). \tag{S8.7}$$

Together with Eq. (S8.1), this proves $G_2^\star(P) = \Theta(P^2)$ under the stated exit cap and memory only readout, rather than only an upper bound.

## S9. Three states attain the logarithmic bound

The three-state design in Figure 1(a) of the main text uses a stale negative state A, a fresh positive state B, and a stale positive state C. Its transition channels depend on the held sensor value as shown in Table S1. The leakage parameter is chosen proportional to the observation frequency:

$$\varepsilon = af, \qquad a = \frac{c}{\gamma}. \tag{S9.1}$$

This construction requires retuning the reverse rates as observations become rare.

**Table S1.** Active transitions in the three-state construction.

| **Held sensor $Z$** | **Forward transition** | **Forward rate** | **Reverse rate** |
|---|---|---|---|
| $+1$ | $A \to B$ | $u$ | $\varepsilon u$ |
| $+1$ | $B \to C$ | $d$ | $\varepsilon d$ |
| $-1$ | $C \to A$ | $u$ | $\varepsilon u$ |
| $-1$ | $B \to A$ | $d$ | $\varepsilon d$ |

For each fixed sensor value the transition network is a reversible tree, so its steady state housekeeping dissipation vanishes. In state order $(A, B, C)$ the frozen sensor-conditioned stationary distributions are

$$\pi_{+1} = \frac{(\varepsilon^2, \varepsilon, 1)}{1 + \varepsilon + \varepsilon^2}, \qquad \pi_{-1} = \frac{(1, \varepsilon, \varepsilon)}{1 + 2\varepsilon}. \tag{S9.2}$$

**Probability and environmental bias of B.** During a long positive hold, the memory remains predominantly in C, and during a negative hold it remains in A. The background occupation of B is $O(\varepsilon)$. Negative to positive sensor changes occur at frequency $f/4 + O(f^2)$ and initiate the excursion $A \to B \to C$. For fixed forward rates $u$ and $d$, the probability that B is occupied at time $t$ after such a change is

$$g_B(t) = \frac{u}{d-u}\left(e^{-ut} - e^{-dt}\right). \tag{S9.3}$$

Its integrated residence time and its overlap with the environmental correlation are

$$\int_0^\infty g_B\,(t)\,dt = \frac{1}{d}, \qquad \int_0^\infty e^{-\lambda t}\, g_B(t)\,dt = \frac{u}{(u+\lambda)(d+\lambda)}. \tag{S9.4}$$

The environmental bias just after a rare negative to positive reading is $\eta$ to leading order and decays as $e^{-\lambda t}$. Adding the background and the event induced occupancy gives

$$p_B = f\left(a + \frac{1}{4d}\right) + o(f), \qquad c_B = \frac{f\eta u}{4(u+\lambda)(d+\lambda)} + o(f). \tag{S9.5}$$

A and C have order one probabilities but environmental biases of order $f$, so their contributions to $G_\tau$ are only $O(f^2)$. The transient B state supplies the leading contribution,

$$G_\tau = K_3 f + o(f), \qquad K_3 = \left(a + \frac{1}{4d}\right)\Phi\left(\frac{q\eta u}{4(u+\lambda)(d+\lambda)\big(a + 1/(4d)\big)}\right). \tag{S9.6}$$

**Energy required for reliable recording.** A complete negative to positive and back switching cycle crosses three driven edges, $A \to B$, $B \to C$, and $C \to A$. Such cycles occur at rate $f/4 + O(f^2)$. The net work per forward edge is $k_B T \ln(1/\varepsilon)$, while leakage excursions typically retrace an edge and make no leading net contribution. Thus

$$P_m = \frac{3}{4} k_B T f \ln(1/\varepsilon) + o\big(f \ln(1/f)\big). \tag{S9.7}$$

Equations (S9.6) and (S9.7) are the two asymptotic statements combined in Eq. (8) of the main text. Since $\varepsilon \propto f$, the construction achieves $G_\tau = \Theta\big(P/\ln(P_{\text{ref}}/P)\big)$ and establishes the lower half of the $n \geq 3$ relation in Eq. (1). The logarithm comes from requiring the background occupancy of B to fall with $f$ without slowing its response to a genuine sensor change.

For $\gamma = 1$, $k_B T = 1$, $\eta = 0.5$, $\tau = 0.4$, $u = 7$, $d = 3$, and $c = 0.01$, Eq. (S9.6) gives $K_3 = 4.095 \times 10^{-4}$. Exact stationary solutions of the twelve-state $(Y, Z, M)$ process give $G_\tau/f = 4.095 \times 10^{-4}$ at $f = 10^{-12}$ and $4.095 \times 10^{-4}$ at $f = 10^{-6}$. They also give $P_m/[k_B T f \ln(1/\varepsilon)] = 0.7500$ for $f \leq 10^{-5}$.

**Additional states.** Splitting a state into identical substates, distributing incoming rates with fixed weights, and retaining outgoing rates preserves the lumped dynamics, predictive information, and exit-rate cap. The additional contributions to work cancel in stationarity. Thus, the asymptotic order is attainable for every $n \geq 3$. Extra states can distribute the affinity over several smaller steps but cannot eliminate the $k_B T\ln(1/f)$ work required to suppress false records.

## S10. Nonadiabatic dissipation and alternative physical implementations

For a fixed generator $W_z$ with positive stationary distribution $\pi_z$, the entropy production separates into nonnegative housekeeping and nonadiabatic contributions [2]. If $p_t$ is the memory distribution conditioned on the sensor history, its evolution and nonadiabatic entropy production during each hold satisfy

$$\dot{p}_t = p_t W_z, \qquad \sigma_{\text{na}} = -\frac{d}{dt} D(p_t \| \pi_z) \geq 0. \tag{S10.1}$$

At a sensor change, $p_t$ is continuous but the reference distribution switches. The exact stationary average is

$$\overline{\sigma}_{\text{na}} = \nu \, \langle D(p_- \| \pi_{z'}) - D(p_- \| \pi_z) \rangle_{\text{flip}}. \tag{S10.2}$$

Equivalently,

$$\overline{\sigma}_{\text{na}} = \sum_{z,i} j_z\,(i) \ln \left[ \frac{\pi_z(i)}{\pi_{-z}(i)} \right]. \tag{S10.3}$$

Here $j_z(i)$ is the sensor flip flux from Section S4, and $\langle \cdot \rangle_{\text{flip}}$ denotes an average over sensor change events weighted by their rates. For constant intrinsic energies, $\beta P_m = \overline{\sigma}_{\text{hk}} + \overline{\sigma}_{\text{na}} \geq 0$.

If the memory fully relaxes between changes *and* the logarithmic weights are controlled, the nonadiabatic cost simplifies to $\nu[D(\pi_{-1} \| \pi_{+1}) + D(\pi_{+1} \| \pi_{-1})]/2$, or $\nu D(\pi_{+1} \| \pi_{-1})$ for sign symmetric memories. This intuitive form is not uniform when relaxation gaps vanish.

The transition rates of a physical memory can also be realized by changing an energy landscape rather than by chemically gating channels. If $U_z(i) = -k_B T \ln \pi_z(i)$ changes when the sensor flips, chemical work can be absent, but the sensor performs switching work $U_{z'}(M) - U_z(M)$. In an implementation without housekeeping dissipation, its average power equals $k_B T \overline{\sigma}_{\text{na}}$. This work must be included in the total thermodynamic accounting: the logarithmic penalty is not specific to chemical fuel.

## S11. Domain of validity

**Free observations and fixed sampling rate.** The bound (S7.8) assumes $E_s > 0$, so that $P \to 0$ implies $f \to 0$. For $E_s = 0$, a reactive two-state memory at fixed $f$ by reducing its chemical affinity $F \to 0$. With $r_0 = 2u$, we define

$$A = \frac{\eta f r_0}{(f + \lambda)(r_0 + \lambda)}, \qquad B = 1 - \frac{\eta^2 f r_0}{(f + \lambda)(r_0 + \lambda)} > 0. \tag{S11.1}$$

The exact expansion is

$$G_\tau = \frac{q^2 A^2}{8} F^2 + o(F^2), \qquad P_m = k_B T \frac{f r_0 B}{4(f + r_0)} F^2 + o(F^2). \tag{S11.2}$$

Thus $G_\tau / P_m$ tends to a positive constant along this different limit: information can grow linearly with chemical power while the fraction of the available sensory information tends to zero.

**Counting all sources of work.** Realizing the three-state network by switching physical energy levels and counting only $fE_s$ would make the predictive information appear proportional to power. The missing switching work is $3/4\, k_B T f \ln(1/\varepsilon)$ at leading order, precisely the recording cost in Eq. (S9.7). Restoring that contribution reinstates the logarithmic dependence.

**Limitations of the approach.** The autonomous memory is driven only by the held sensor value. It does not receive observation timestamps, synchronized reset pulses, an external clock, an explicitly tracked age variable, or a free precharged record. The analysis does not cover feedback from memory to sensor or general active, inertial, quantum, or continuous state memories. For $\eta < 1$, the universal power bound can be proved without an exit cap by the modification noted in Section S5. The two-state quadratic theorem uses the cap at every $\eta$. At perfect accuracy $\eta = 1$, the cap is needed for the uniform theorem. Whether it can be removed in that case, and whether the two-state quadratic bound survives without it, remain open questions.

## S12. Numerical methods and checks

**Stationary distribution.** For $n$ internal memory states, the complete $(Y, Z, M)$ process has $4n$ states. Its stationary distribution was computed with the Grassmann–Taksar–Heyman state elimination algorithm [3], which preserves positivity when transition rates span many orders of magnitude. The predictive information and the channel resolved chemical power were then evaluated from Eqs. (S1.6) and (S1.3). For the four-state curve in Fig. 2, the normalized 16-state stationary linear system was also solved directly, and each accepted solution was checked by its stationarity residual.

**Random networks.** As a consistency check, we tested 400 random networks with $n = 2,3,4,6$, $10^{-6} \leq f/\gamma \leq 0.3$, $0.05 \leq \eta \leq 0.999$, $0.1 \leq \kappa/\gamma \leq 100$, and $0 \leq \gamma\tau \leq 1$. The rates were sampled log uniformly and rescaled to respect the exit cap. No violation was found of the posterior bias bound (S3.1), entropy bound (S5.3), finite sampling bound (S7.2), sensory benchmark, or the two-state theorem. The largest observed ratios of the left to right sides were respectively 0.94, 0.99, 0.0097, 0.17, and 0.0063. In particular, the universal finite sampling coefficients and the two-state coefficients are conservative. The random tests are consistency checks, not demonstrations of sharpness.

**Optimized designs and Figure 2.** At fixed total power budget, the two-state reactive family was searched deterministically over its kinetic rate and affinity, whereas the three-state line is the explicit (unoptimized) construction of Section S9. The four-state curve uses a sign-symmetric recency architecture with A (stale negative), B (fresh positive), C (stale positive), and D (fresh negative). For a positive held sensor, its reversible channels are A↔B (record), B↔C (age), and D↔C (reset); for a negative sensor they become C↔D, D↔A, and B↔A. Their six forward and reverse rates are optimized independently without prescribing a sampling-dependent leakage law.

The sampling frequency and six positive rates were varied to maximize predictive information subject to the total power budget $P = fE_{\mathrm{s}} + P_{\mathrm{m}}$ and the maximum memory escape rate $\kappa = 10\gamma$. The plotted search used deterministic multistart sequential quadratic programming and continuation in both directions along the power grid; the numerical lower rate floor is independent of $f$. The 16-state stationary distribution, channel-resolved chemical work, stationary residual, and physical constraints were checked for each accepted solution. Each fixed-sensor network is a tree and therefore has no stationary housekeeping cycle. Only verified feasible values are connected, with no smoothing. The four-state curve is a best-found restricted-topology construction, not a certified global optimum or a proof of asymptotic scaling. Its advantage is relative to the particular unoptimized three-state reference. The gray line represents the sensor change history information bound Eq. (S2.9), not a thermodynamic work bound.

**Sensor change filter.** The process $(Y, Z)$ was simulated exactly using the Gillespie algorithm [4]. Between changes, the Riccati equation (S2.6) was integrated analytically and time averages were accumulated by Gauss–Legendre quadrature over each interval, using approximately $1.2 \times 10^4$ to $2 \times 10^4$ trajectories. Independent runs agree at the numerical precision reported in Section S2.